\documentclass{article}

\usepackage[utf8]{inputenc}
\usepackage[margin=1.25in]{geometry} 
\usepackage[hidelinks]{hyperref} 
\usepackage{setspace} 
\usepackage{float} 
\usepackage{graphicx} 
\usepackage[labelfont=bf]{caption} 

\usepackage{changepage}

\usepackage{amsmath}
\usepackage{amssymb} 
\usepackage{bm}
\usepackage{makecell} 

\newlength\savedwidth

\newcommand\thickhline{\noalign{\global\savedwidth\arrayrulewidth\global\arrayrulewidth 1pt}%
\hline
\noalign{\global\arrayrulewidth\savedwidth}}

\begin{document}

\begin{flushleft}

\begin{spacing}{1.6}
{\LARGE \textbf{RIDDLE: Race and ethnicity Imputation from Disease history with Deep LEarning}}
\end{spacing}

Ji-Sung Kim\textsuperscript{\normalfont{1}}, Xin Gao\textsuperscript{\normalfont{2}}, Andrey Rzhetsky\textsuperscript{\normalfont{3*}} \\

\bigskip

\textbf{1} Center for Statistics and Machine Learning, Department of Computer Science, Princeton University, Princeton, New Jersey, United States of America \\
\textbf{2} King Abdullah University of Science and Technology (KAUST), Computational Bioscience Research Center (CBRC), Computer, Electrical and Mathematical Sciences and Engineering (CEMSE) Division, Thuwal, Saudi Arabia. \\
\textbf{3} Institute for Genomics and Systems Biology,
Computation Institute,
Departments of Medicine and Human Genetics, University of Chicago, Chicago, Illinois, United States of America \\

\bigskip

* andrey.rzhetsky@uchicago.edu

\rule{\textwidth}{0.5pt}

\end{flushleft}

\textbf{This manuscript was revised and published in \textit{PLOS Computational Biology}. This arXiv preprint is now outdated; the updated article is available as a free and open access publication at  \href{https://doi.org/10.1371/journal.pcbi.1006106}{https://doi.org/10.1371/journal.pcbi.1006106}. }

\section*{Abstract}

Anonymized electronic medical records are an increasingly popular source of research data. However, these datasets often lack race and ethnicity information. This creates problems for researchers modeling human disease, as race and ethnicity are powerful confounders for many health exposures and treatment outcomes; race and ethnicity are closely linked to population-specific genetic variation. We showed that deep neural networks generate more accurate estimates for missing racial and ethnic information than competing methods (e.g., logistic regression, random forest). RIDDLE yielded significantly better classification performance across all metrics that were considered: accuracy, cross-entropy loss (error), and area under the curve for receiver operating characteristic plots (all $p < 10^{-6}$). We made specific efforts to interpret the trained neural network models to identify, quantify, and visualize medical features which are predictive of race and ethnicity. We used these characterizations of informative features to perform a systematic comparison of differential disease patterns by race and ethnicity. The fact that clinical histories are informative for imputing race and ethnicity could reflect (1) a skewed distribution of blue- and white-collar professions across racial and ethnic groups, (2) uneven accessibility and subjective importance of prophylactic health, (3) possible variation in lifestyle, such as dietary habits, and (4) differences in background genetic variation which predispose to diseases.

\section{Introduction}
Electronic medical records (EMRs) are an increasingly popular source of biomedical research data~\cite{jensen2012mining}. EMRs are digital records of patient medical histories, describing the occurrence of specific diseases and medical events such as the observation of heart disease or dietary counseling. EMRs can also contain demographic information such as gender or age.

However, these datasets are often anonymized and lack race and ethnicity information (e.g., insurance claims datasets). Race and ethnicity information may also be missing for specific individuals within datasets. This is problematic in research settings as race and ethnicity can be a powerful confounder for a variety of effects. Race and ethnicity are strong correlates of socioeconomic status, a predictor of access to and quality of education and healthcare. These factors are differentially associated with disease incidence and trajectories. As a result of this correlation, race and ethnicity may be associated with variation in medical histories. As an example, it has been reported that referrals for cardiac catheterization are rarer among African American patients than in White patients~\cite{schulman1999effect}. Furthermore, researchers have reported differences in genetic variation which influence disease across racial and ethnic groups~\cite{burchard2003importance}. Due to the association between race, ethnicity and medical histories, we hypothesize that clinical features in EMRs can be used to impute missing race and ethnicity information.

In addition, race and ethnicity information can be useful for producing and investigating hypotheses in epidemiology. For example, variation in disease risk across racial and ethnic groups that cannot be fully explained by allele frequency information may provide insights into the possible environmental modifiers of genes~\cite{burchard2003importance}.

\subsection{Imputation}

The task of race and ethnicity imputation can be serialized as a supervised learning problem. Typically, the goal of imputation is estimate a posterior probability distribution over plausible values for a missing variable. This distribution of plausible values can be used to generate a single imputed dataset (e.g., by choosing plausible values with highest probability), or to generate multiple imputed datasets as in \emph{multiple imputation}~\cite{sterne2009multiple}. In our setting, the goal was to impute the distribution of mutually-exclusive racial and ethnic classes given a set of clinical features. Features comprised age, gender, and codes from the International Disease Classification, version 9 (ICD9, \cite{RN25809}); ICD9 codes describe medical conditions, medical procedures, family information, and some treatment outcomes.

Bayesian approaches to race and ethnicity imputation using census data have been proposed~\cite{elliott2008new} and have been used for race and ethnicity imputation in EMR datasets~\cite{grundmeier2015imputing}. However, these approaches require sensitive geolocation and surname data from patients. Geolocation and surname data can be missing in anonymized EMR datasets (as in the datasets used here), limiting the utility of approaches which use this information.


\subsection{Deep learning}

Traditionally, logistic regression classifiers have been used to impute categorical variables such as race and ethnicity~\cite{sentas2006categorical}. However, there has been recent interest in the use of deep learning for solving similar supervised learning tasks. Deep learning is particularly exciting as it offers the ability to automatically learn complex representations of high-dimensional data. These representations can be used to solve learning tasks such as regression or classification~\cite{lecun2015deep}.

Deep learning involves the approximation of some utility function (e.g., classification of an image) as a neural network. A neural network is a directed graph of functions which are referred to as units, neurons or nodes. This network is organized into several layers; each layer corresponds to a different representation of the input data. As the input data is transformed and propagated through this network, the data at each layer corresponds to a new alternate-dimension representation of the sample~\cite{lecun2015deep}. For our imputation task, the aim was to learn the representation of an individual as a mixture of race and ethnicity classes where each class is assigned a probability. This representation is encoded in the final output layer of the neural network. The output of a neural network functions as a prediction of the distribution of race and ethnicity classes given a set of input features.

\begin{figure}[H]
\begin{center}
\includegraphics[width=0.95\textwidth]{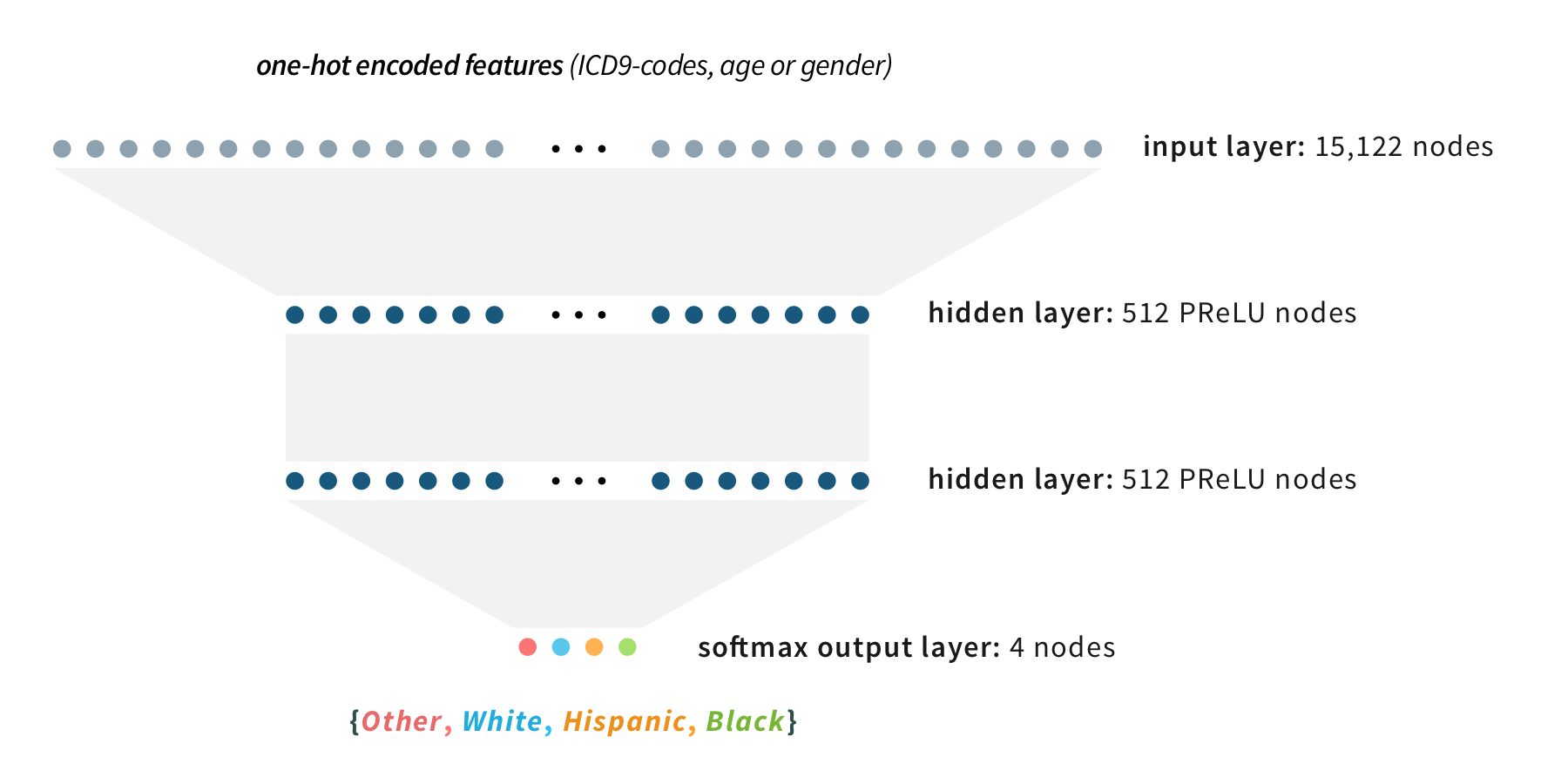}
\end{center}
\caption{{\bf Neural network architecture.} RIDDLE uses a multi-layer perceptron (MLP) network containing two hidden layers of Parametric Rectified Linear Unit (PReLU) nodes. The input to the MLP is the set of binary encoded features comprising age, gender, and International Disease Classification, version 9 (ICD9) codes. The output is the set of probability estimates for each of the four race and ethnicity classes. }
\label{fig1}
\end{figure}

We introduce a framework for using deep learning to estimate missing race and ethnicity information in EMR datasets: \textbf{RIDDLE} or \textbf{R}ace and ethnicity \textbf{I}mputation from \textbf{D}isease history with \textbf{D}eep \textbf{LE}arning. RIDDLE uses a relatively simple multilayer perceptron (MLP), a type of neural network architecture that is a directed acyclic graph (see Fig~\ref{fig1}). For its nodes, our neural network architecture utilizes Parametric Rectifier Linear Units (PReLUs)~\cite{he2015delving}, which are rectifier functions:
$$f(x) =
\begin{cases}
      x, &  x > 0; \\
      \alpha x, & x \leq 0,
\end{cases}
\quad\quad \text{where } \alpha \text{ is a learned parameter.}
$$
where $x$ is the input, and $f(x)$ is the output of the PReLU node. Further details of RIDDLE's implementation are described in the~\hyperref[methods]{\emph{Methods}}.

In addition to investigating the novel utility of deep learning for race and ethnicity imputation, we used recent methods in interpreting neural network models~\cite{shrikumar2016not} to perform a systematic evaluation of racial and ethnic patterns for approximately 15,000 different medical events. We believe that this type of large-scale evaluation of disease patterns and maladies by race and ethnicity has not been done heretofore.

\section{Results}


We aimed to assess RIDDLE's imputation performance in a multiclass classification setting. We used EMR datasets from Chicago and New York City, collectively describing over 1.5 million unique patients. There were approximately 15,000 unique input features consisting of basic demographic information (gender, age) and observations of clinical events (codified as ICD9 codes). The target class was race and ethnicity; possible values were White, Black, Other or Hispanic. Although race and ethnicity can be described as a mixture, our training datasets labeled race and ethnicity as one of four mutually exclusive classes. For the testing set, we treated the target race and ethnicity class as missing, and compared the predicted class against the true class. The large dimensionality of features, high number of samples, and heterogeneity of the source populations present a unique and challenging classification problem.

\begin{figure}[H]
\begin{center}
\includegraphics[width=1.0\textwidth]{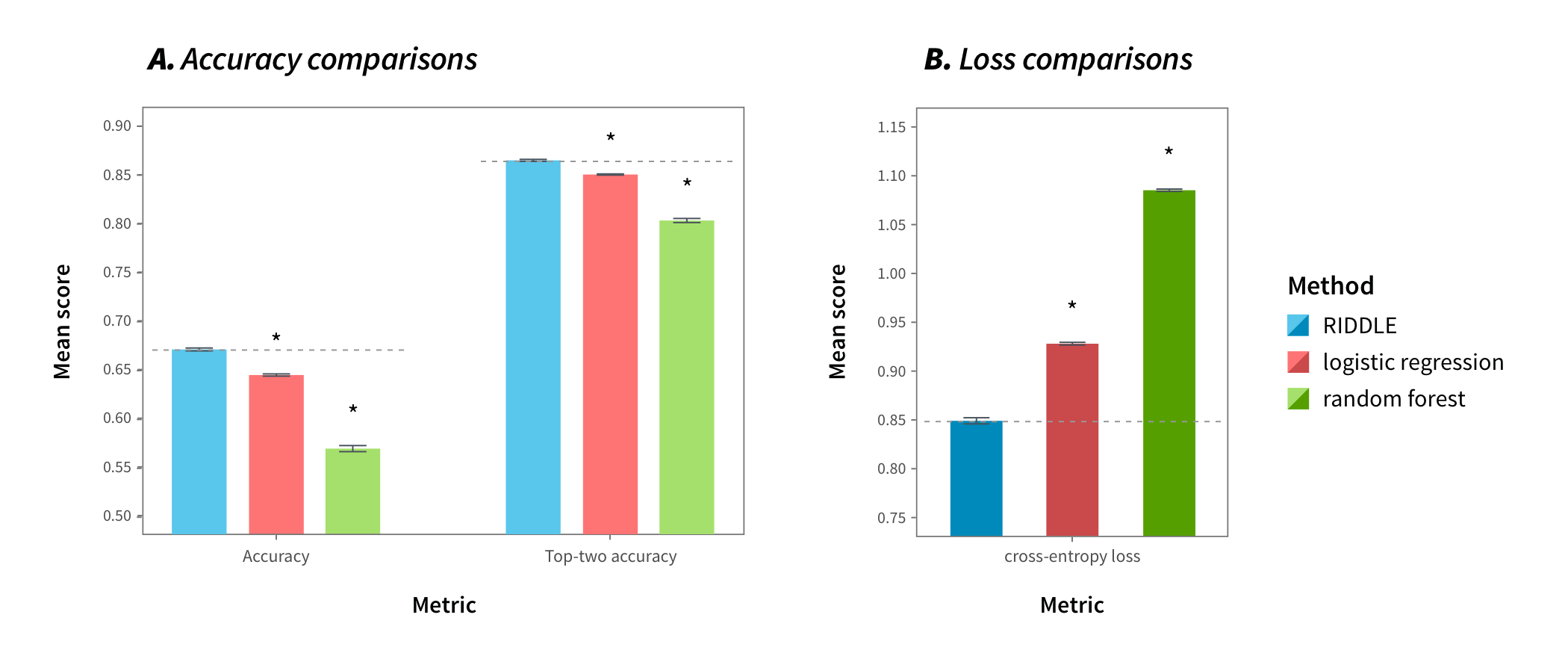}
\end{center}
\caption{{\bf Accuracy and loss comparisons of RIDDLE and other methods.}
RIDDLE outperformed popular classification methods, achieving significantly higher accuracy ($p=1.33 \times 10^{-10}$, $2.63 \times10^{-14}$ compared to logistic regression, random forest respectively), higher top-two accuracy ($p=2.10 \times 10^{-10}$, $1.26 \times 10^{-14}$), and lower cross-entropy loss ($p=7.52 \times 10^{-13}$, $2.29 \times 10^{-16}$). Higher accuracy and lower loss indicate better performance. Scores were averaged over ten $k$-fold cross-validation experiments. }
\label{fig2}
\end{figure}

In our experiments, RIDDLE yielded an average accuracy of 0.671, top-two accuracy of 0.865, and cross-entropy loss of 0.849 on test data, significantly outperforming logistic regression and random forest classifiers (see Fig~\ref{fig2}). Support vector machines (SVMs) with various kernels were also evaluated. However, SVMs could not be feasibly used as the computational cost was too high; each experiment required more than 36 hours.

\begin{figure}[H]
\begin{center}
\includegraphics[width=0.8\textwidth]{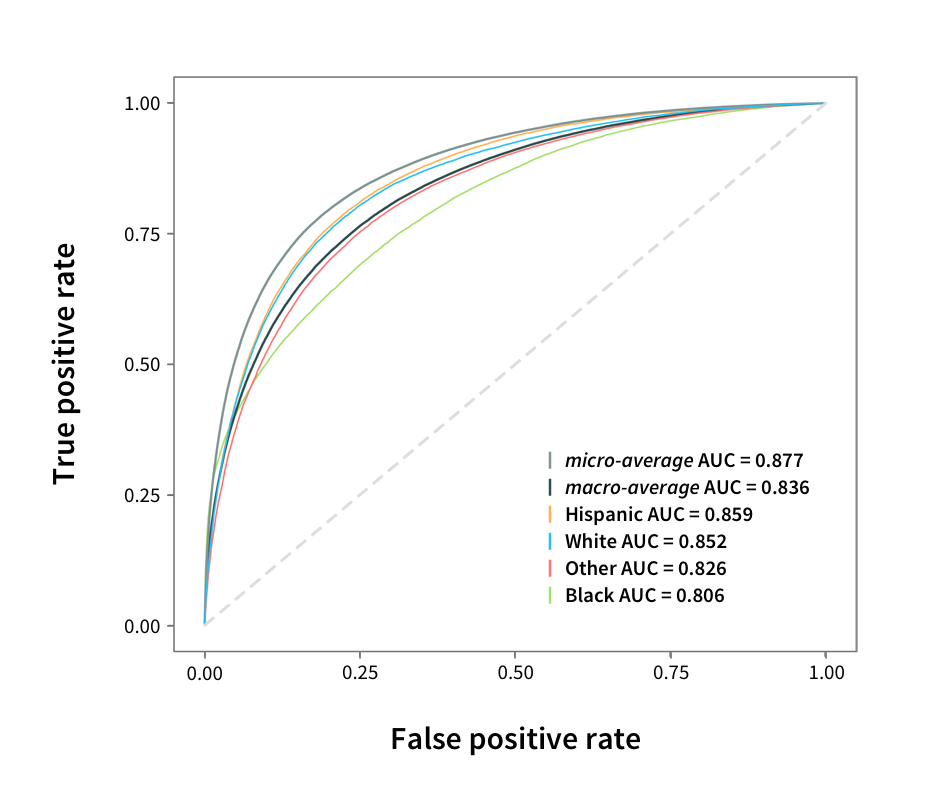}
\end{center}
\caption{{\bf Receiver operating characteristic (ROC) curves.}
ROC curves and their corresponding area under the curve (AUC) values were calculated for each of the four race and ethnicity classes. Micro-average (all cases considered as binary, e.g., Hispanic vs. non-Hispanic) and macro-average (average across classes) curves were also computed. Data for a representative experiment is shown. Across experiments, the mean micro-average AUC was 0.877, and the mean macro-average AUC was 0.836.}
\label{fig3}
\end{figure}

While the multiclass learning problem appeared relatively hard, RIDDLE exhibited class-specific receiver operating characteristic's (ROC) area under the curve (AUC) values above 0.8 (see Fig~\ref{fig3}), and a mean micro-average (all cases considered as binary) AUC of 0.877 -- significantly higher than that of logistic regression (mean=0.854, $p=1.45 \times 10^{-13}$) and random forest (mean=0.799, $p=1.23\times 10^{-16}$) classifiers.



RIDDLE exhibited runtime performance comparable to that of other machine learning methods on a standard computing configuration without the use of a graphics processing unit or GPU (see Table~\ref{table1}). Support vector machines were also evaluated, but precise runtime measurements could not be obtained as experiments took greater than 36 hours each (36 hours runtime was the allowed maximum on the system used in our analysis). However, on a smaller subset (150K samples) of the full dataset, RIDDLE exhibited significantly better classification accuracy and faster runtime performance than SVMs with various kernels (see Table~\ref{table3} in the \text{Supporting Information}).

\begin{table}[H]
\centering
\begin{tabular}{ l c}
\thickhline
Method & Average runtime $\pm \text{ } SD$ (h) \\
\hline
\\[-0.25cm]
RIDDLE                 & $1.318 \pm 0.165$ \\
logistic regression    & $0.376 \pm 0.002$ \\
random forest          & $1.429 \pm 0.034$ \\
SVM, linear kernel     & \textgreater $36$ \\
SVM, polynomial kernel & \textgreater $36$ \\
SVM, RBF kernel        & \textgreater $36$ \\
\thickhline
\\[-0.2cm]
\end{tabular}
\caption{{\bf Runtime comparisons of RIDDLE and other methods.}
RIDDLE demonstrated runtimes (combined training and testing time) between that of logistic regression and random forest classifiers. Runtime values were averaged over ten $k$-fold cross-validation experiments. A standard computing configuration was used: 16 Intel Sandybridge cores at 2.6 GHz; graphics processing units were not utilized. }
\label{table1}
\end{table}

\subsection{Influence of missing data on classifier performance}

\begin{figure}[H]
\begin{center}
\includegraphics[width=0.85\textwidth]{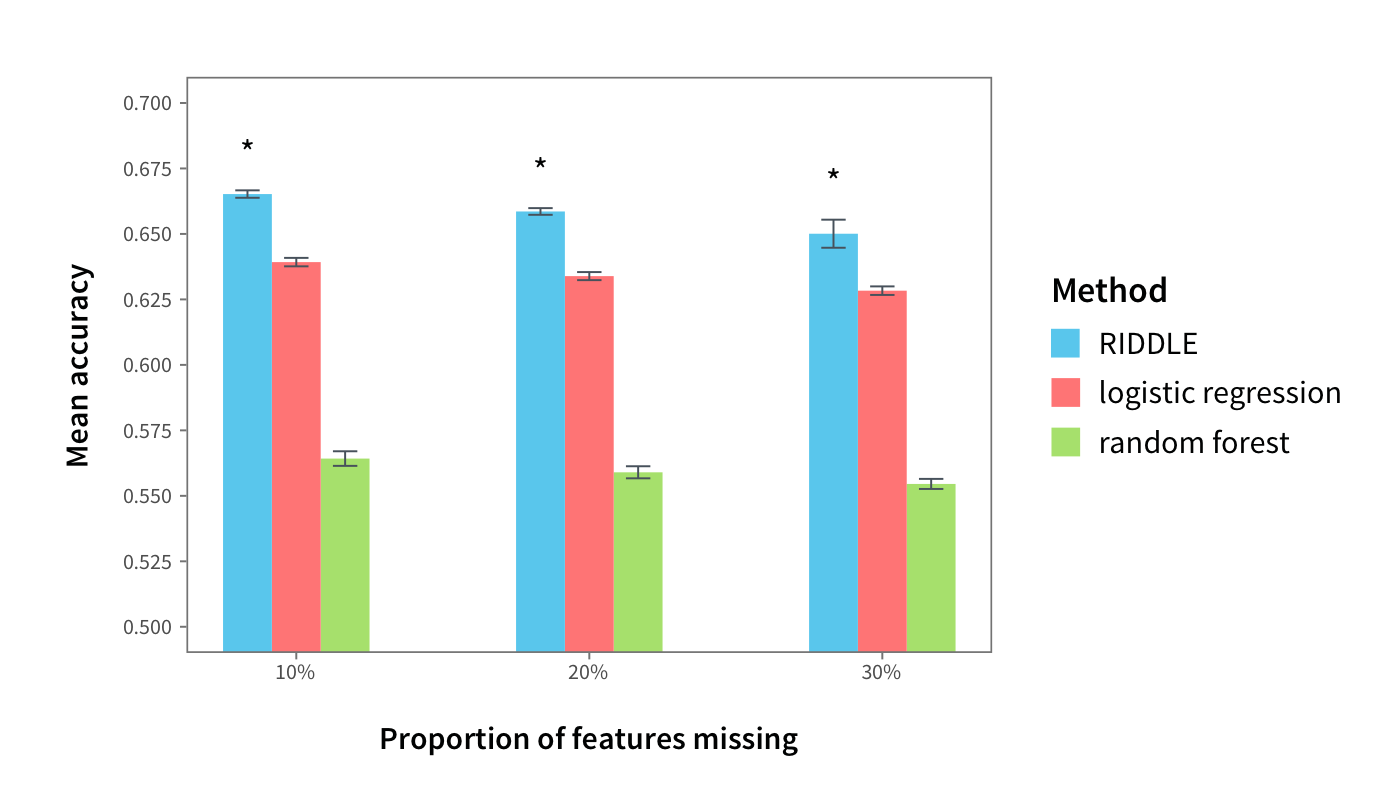}
\end{center}
\caption{{\bf Accuracy of RIDDLE method compared to standard techniques in situations with missing data.} In experiments simulating different proportions of random missing data (10\%-30\%), RIDDLE outperformed logistic regression and random forest classifiers. RIDDLE yielded significantly higher accuracy scores: in the 10\% missing data experiment, $p=3.78 \times 10^{-14}, 5.08 \times 10^{-16}$ for RIDDLE compared to logistic regression and random forest respectively; in the 20\% missing data experiment, $p=2.38 \times 10^{-13}, 1.40 \times 10^{-15}$; in the 30\% missing data experiment, $p=5.17 \times 10^{-7}, 1.05 \times10^{-12}$. Accuracy scores were averaged over ten $k$-fold cross-validation experiments.
}
\label{fig4}
\end{figure}

In order to replicate real-world applications where data other than race and ethnicity (e.g., features) may be missing, we conducted additional experiments to simulate random missing data. A random subset of \textit{sample} features (ranging from 10\% to 30\% of all features) was artificially masked completely at random. Otherwise, the same classification training and evaluation scheme was used as before. Under simulation of random missing data, RIDDLE significantly outperformed logistic regression and random forest classifiers in terms of classification accuracy (see Fig~\ref{fig4}).

\subsection{Feature interpretation}

A major criticism of deep learning is the opaqueness of trained neural network models for intuitive interpretation. While intricate functional architectures enable neural networks to learn complex tasks, they also create a barrier to understanding how learning decisions (e.g., classifications) are made.
In addition to creating a precise race and ethnicity estimation framework, we sought to identify and describe the factors which contribute to these estimations. We computed DeepLIFT (Learning Important FeaTures) scores to quantitatively describe how specific features contribute to the probability estimates of each class. The DeepLIFT algorithm compares the activation of each node to a reference activation; the difference between the reference and observed activation is used to compute the contribution score of a neuron to a class~\cite{shrikumar2016not}.

If a feature contributes to selecting \emph{for} a particular class, this feature-class pair is assigned a positive DeepLIFT score; conversely, if a feature contributes to selecting \emph{against} a particular class, the pair is assigned a negative score. The magnitude of a DeepLIFT score represents the strength of the contribution.

Using DeepLIFT scores, we were able to construct natural orderings of race and ethnicity classes for each feature, sorting classes by positive to negative scores. The following example ordering shows how the example feature (heart disease) is a strong predictor for the African American class, and a weak (or negative) predictor for the Other class.
$$
\begin{aligned}
      \text{heart disease} \to \text{Other, } &\text{score}=-500 \\
      \text{heart disease} \to \text{White, } &\text{score}=-100 \\
      \text{heart disease} \to \text{Hispanic, } &\text{score}=+200 \\
      \text{heart disease} \to \text{Black, } &\text{score}=+500 \\
\end{aligned}
\quad\Longrightarrow\quad
\text{Black} > \text{Hispanic} > \text{White} > \text{Other}
$$

We computed the class orderings for all $\sim$15,000 features. The orderings of the 10 most predictive features (by highest ranges of DeepLIFT scores) are described in Table~\ref{table2}.

\bgroup
\def\arraystretch{1.25}
\begin{table}[H]
\centering
\begin{adjustwidth}{-0.5in}{0in}
\begin{tabular}{l l p{2in} c}
\thickhline
Rank & ICD9 & Description & Ordering of race and ethnicity classes (DeepLIFT scores) \\
\hline
1 & V72.6 & Laboratory examination & W (95780) $>$ O (61079) $>$ B (-44595) $>$ H (-112264) \\
2 & 401.9 & Hypertension NOS & B (49720) $>$ H (43333) $>$ O (-46417) $>$ W (-46635) \\
3 & 789.00 & Abdominal pain, unspecified site & H (24354) $>$ B (12191) $>$ W (-5949) $>$ O (-30595) \\
4 & V65.44 & Human immunodeficiency virus [HIV] counseling & H (29684) $>$ B (12788) $>$ O (-17647) $>$ W (-24825) \\
5 & V20.2 & Routine infant or child health check & H (32848) $>$ B (4931) $>$ W (-18495) $>$ O (-19284) \\
6 & V70.0 & Routine general medical examination at a health care facility & H (29286) $>$ B (946) $>$ W (-8077) $>$ O (-22155) \\
7 & V30.00 & Single liveborn, born in hospital, delivered without mention of cesarean section & W (26671) $>$ B (4246) $>$ H (-8100) $>$ O (-22816) \\
8 & V72.9 & Unspecified examination & W (33491) $>$ B (-5071) $>$ H (-13956) $>$ O (-14464) \\
9 & 465.9 & Acute upper respiratory infections of unspecified site & B (17540) $>$ H (13466) $>$ O (-8496) $>$ W (-22510) \\
10 & 079.9 & Unspecified viral and chlamydial infection in conditions classified elsewhere and of unspecified site & O (17789) $>$ B (4851) $>$ H (-4351) $>$ W (-18290) \\
\thickhline
\\[-0.2cm]
\end{tabular}
\end{adjustwidth}
\caption{{\bf DeepLIFT contribution score orderings for 10 most \emph{predictive} ICD9 codes.}
DeepLIFT scores were computed for each pair of feature, and race and ethnicity class; we list ten ICD9 codes with the highest ranges of scores -- which correspond to discriminative ability. The feature-to-class contribution scores were used to construct orderings of race and ethnicity classes, for each feature. DeepLIFT scores were summed across all samples. Positive scores indicate favorable contribution to a class, zero scores indicate no contribution, and negative scores indicate discrimination against a class. }
\label{table2}
\end{table}
\egroup

\begin{figure}[H]
\begin{center}
\includegraphics[width=0.72\textwidth]{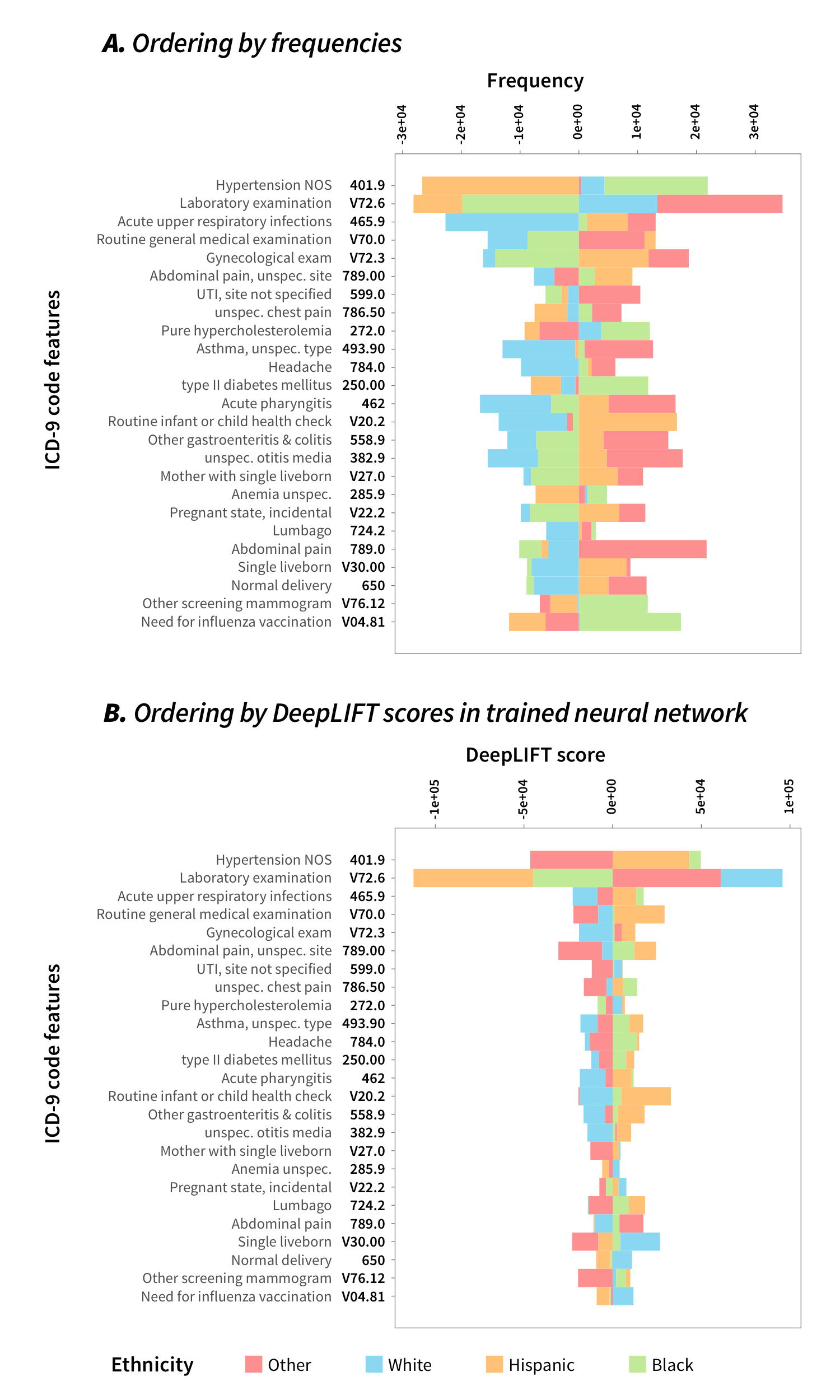}
\end{center}
\caption{{\bf Visualizing class orderings for the 25 most \emph{common} features.}  We constructed natural orderings of features for the 25 most common ICD9 code features, using (A) frequency information and (B) DeepLIFT scores. Frequency scores were mean-centered; higher scores indicate larger contribution by a feature to a class. These orderings rank the contribution of an ICD9 code to a particular class, and are visualized as a stacked bar. The strongest (positive) feature-to-class contributions are represented by the rightmost bar; the length of the bar corresponds to the magnitude of the contribution on a linear scale. Scores were summed across all samples. }
\label{fig5}
\end{figure}

We visualized the orderings of the 25 most \emph{common} features using both frequencies and DeepLIFT scores (see Fig~\ref{fig5}). Race and ethnicity class orderings obtained from frequency scores were distinctly different than those obtained from DeepLIFT scores. This suggests that RIDDLE's MLP network is able to learn non-linear and non-frequentist relationships between ICD9 codes and race and ethnicity categories.

According to orderings constructed using DeepLIFT scores, sex is an important feature for predicting race and ethnicity in our models: men who seek medical attention are least likely to be African American, followed by Hispanic men. Men who seek medical attention are most likely to be White or Other.

In addition, specific medical diagnoses convey grains of racial and ethnic information: hypertension and human immunodeficiency virus (HIV) are more predictive for African American individuals than White individuals. This finding is also reflected in medical literature, where it has been reported that African Americans are at significantly higher risk for heart disease~\cite{RN35850,RN35852} and HIV~\cite{crepaz2007efficacy,RN35881} than their White peers.


The fact that these features are important for imputing race and ethnicity could reflect (1) a skewed distribution of blue- and white-collar professions across racial and ethnic groups, (2) uneven accessibility and subjective importance of prophylactic health care across racial and ethnic groups, (3) and possible variation in lifestyle, such as dietary habits. Further work would involve investigating epidemiological hypotheses on how these environmental factors may affect differential clinical patterns across race and ethnicity.

Some of the genetic diseases are famously discriminative for races and ethnicities. For example, sickle cell disease occurs 88 time more frequently in African Americans than in the rest of the US population \cite{RN35809}. In our model, sickle cell anemia most strongly predicts for the African American race. It has been reported Lyme disease predominately occurs in Whites, and largely unreported for Hispanics or African Americans ~\cite{fix2000racial}. This finding is also reflected in our model, where Lyme disease serves as a strong predictor of the White race. Additional strongly White-predictive diseases and medical procedures include atrial fibrillation, hypothyroidism, prostate neoplasm, dressing and sutures, lump in breast, coronary atherosclerosis. These are primarily diseases of older age, suggesting that lifespan varies across race and ethnicity due to socioeconomic and lifestyle reasons.

These orderings provide a high-level description of community structure, and may reflect socioeconomic, cultural, habitual, and genetic variation linked to race and ethnicity across the population of two large cities, New York City and Chicago.

\section{Discussion}

In our experiments, RIDDLE yielded favorable classification performance with class-specific AUC values of above 0.8. RIDDLE displayed significantly better classification performance across all tested metrics compared to the popular classification methods logistic regression and random forest. RIDDLE's superior top-two accuracy and loss results suggest that RIDDLE produces more accurate probability estimates for race and ethnicity classes compared to currently used techniques. Although results could not be obtained for SVMs due to unacceptably high computational costs, RIDDLE outperformed SVMs in runtime efficiency and classification performance on a smaller subset of the full dataset (see Table~\ref{table3} in the \text{Supporting Information}).

Furthermore, RIDDLE, without the use of a GPU, displayed runtimes comparable to those of traditional classification techniques and required less memory. With these findings, we argue that deep-learning-driven imputation offers notable utility for race and ethnicity imputation in anonymized EMR datasets. Our current work simulated conditions where ethnicity was missing completely at random. Future work will involve simulating conditions where race and ethnicity are missing at random or missing not at random, and formalizing a multiple imputation framework involving deep-learning estimators.

However, these results also highlight a growing privacy concern. It has been shown that the application of machine learning poses non-trivial privacy risks, as sensitive information can be recovered from non-sensitive features~\cite{calandrino2011you}. Our results underscore the need for further anonymization in clinical datasets where race and ethnicity are private information; simple exclusion is not sufficient.

In addition to assessing the predictive and computational performance of our imputation framework, we made efforts to analyze how specific features contribute to race and ethnicity imputations in our neural network model. Each individual feature may represent only a weak trend, but together numerous indicators can synergize to provide a compelling evidence of how a person’s lifestyle, her social circles, and even genetic background can vary by race and ethnicity.

The aforementioned highlights of race- and ethnicity-influenced patterns of health diversity and disparity (see the \textit{Results}) can be extended to thousands of codes. To the best of our knowledge, this systematic comparison across all classes of maladies with respect to race and ethnicity is done for the first time in our study.

\section{Methods} \label{methods}

\subsection{Ethics Statement}

Our study used de-identified, independently collected patient data, and was determined by the Internal Review Board (IRB) of the University of Chicago to be exempt from further IRB review, under the Federal Regulations category 45 CFR 46.101(b).

\subsection{Data}

We used an anonymized EMR datasets jointly comprising 1,650,000 individual medical histories from the New York City (Columbia University) and Chicago metropolitan populations (University of Chicago). Medical histories are encoded as variable length lists of ICD9 codes (approximately 15,000 unique codes) coupled with onset ages in years. Each individual belongs to one of four mutually exclusive classes of race (Other, White, Black) or ethnicity (Hispanic). Features included quinary gender (male, female, trans, other, unknown), and reported age in years.

Onset age information of each ICD9 code was removed and continuous age information was coerced into discrete integer categories. Features were vectorized in a binary encoding scheme, where each individual is represented by a binary vector of zeros (feature absent) and ones (feature present). Each element in the binary encoded vector corresponds to an input node in the trained neural network (see Fig~\ref{fig1}).

$k$-fold cross-validation ($k=10$) and random shuffling were used to produce ten complementary subsets of training and testing data, corresponding to ten classification experiments; this allowed for test coverage of the entire dataset. From the training set, approximately 10\% of samples were used as holdout validation data for parameter tuning and performance monitoring. Testing data was held out separately and was only used during the evaluation process.

\subsection{Hyperparameter tuning}

Hyperparameters were selected using randomized grid search on the 10,000 samples from the validation data. It has been reported that randomized grid search requires far less computational effort than exhaustive grid search with only slightly worse performance~\cite{bengio2012practical}.

\subsection{A deep learning approach}

We used Keras~\cite{chollet2015keras} with a TensorFlow backend~\cite{abadi2016tensorflow} to train a deep multilayer perceptron (MLP) with parameterized rectified linear units (PReLU)~\cite{he2015delving}. The network was composed of an input layer of 15,122 nodes, two hidden layers of 512 PReLU nodes each, and a softmax output layer of four nodes (see Fig~\ref{fig1}). Dropout regularization was applied to each hidden layer~\cite{srivastava2014dropout}.
The MLP was trained iteratively using the \emph{Adam} optimizer~\cite{kingma2014adam}. Training was performed in a batch-wise fashion; data vectorization (via binary encoding) was also done batch-wise in coordination with training. The large number of samples (1.65M) and attention to scalability necessitated ``on the fly`` vectorization. The number of training epochs (passes over the data) was determined by early stopping with patience and model caching~\cite{bengio2012practical}, where the model from the epoch with minimal validation loss was selected.

Categorical cross-entropy was chosen as the loss function; categorical cross-entropy penalizes the assignment of lower probability on the correct class and the assignment of non-zero probability to incorrect classes.

\subsection{Other machine learning approaches}

We evaluated several other machine learning approaches: random forest classifier, logistic regression, and support vector machines (SVMs) with various kernels (linear, polynomial, radial basis function). Traditionally, logistic regression has been used for categorical imputation tasks~\cite{sentas2006categorical}. We used fast Cython (C compiled from Python) or array implementations of these methods offered in the popular `scikit-learn` library.

\subsection{Missing data simulation}
In order to replicate real-world scenarios where additional information (other than race and ethnicity) may be absent, we conducted simulation experiments where we randomly removed some proportion of feature data (10\%, 20\%, or 30\%). We conducted separate training and testing pipelines with these new ``deficient`` datasets.

\subsection{Evaluation}

We computed standard accuracy and cross-entropy loss scores for testing data across all ten experiments. We also computed top-two accuracy, a specification of top-$k$ accuracy. In top-$k$ accuracy, a prediction is considered correct if the true class is contained within the $k$ classes with highest probability assignments.
In addition to evaluating classification performance, we also monitored runtime performance across methods. Models were trained on a standard computing configuration: 16 Intel Sandybridge cores at 2.6 GHz.

Significant differences in performance scores were detected using paired t-tests with Bonferroni adjustment.

\subsection{Neural network interpretation}

We computed DeepLIFT scores to interpret how certain features contribute to probability estimates for each class ~\cite{shrikumar2016not}. The DeepLIFT algorithm takes a trained neural network and produces feature-to-class contribution scores for each passed sample. We computed DeepLIFT scores using test samples in each of our $k$-fold cross validation experiments, to achieve full coverage of the dataset. To describe high-level relationships between features and classes, we summed scores across all samples to produce an aggregate score. The aggregate DeepLIFT scores for the ten most predictive features are summarized in Table~\ref{table2}.
As described prior, we computed orderings of race and ethnicity classes with each feature's DeepLIFT scores. These orderings describe how certain features (e.g., medical conditions) can predict for or against a particular race and ethnicity class. We visualize the orderings defined by DeepLIFT scores for the twenty-five most common features in Fig~\ref{fig5}, and compare them to orderings produced from total frequencies of feature-class observations. For the visualizations, frequency counts were mean-centered to facilitate comparison to DeepLIFT scores.

\section*{Acknowledgments}

We are grateful to Drs. Raul Rabadan and Rachel Melamed for preparing the Columbia University dataset. This work was funded by the DARPA Big Mechanism program under ARO contract W911NF1410333, by NIH grants R01HL122712, 1P50MH094267, U01HL108634-01, and a gift from Liz and Kent Dauten.

\medskip

\bibliographystyle{ieeetr}

\section*{Supporting Information}

\paragraph*{Implementation}
RIDDLE is available as an open-source Python library, \texttt{riddle}. The code is hosted on \href{https://www.github.com/jisungk/riddle}{GitHub}, and documentation is available at \href{https://riddle.ai}{riddle.ai}.

\begin{table}[H]
\centering
\begin{tabular}{ l c c}
\thickhline
Method                 & Average accuracy $\pm \text{ } SD$ & Average runtime $\pm \text{ } SD$ (s) \\
\hline
\\[-0.25cm]
RIDDLE                 & $0.630 \pm 0.009$ & $19.2 \pm 9.9$  \\
logistic regression    & $0.624 \pm 0.013$ & $21.6 \pm 3.5$ \\
random forest          & $0.570 \pm 0.017$ &  $2.7 \pm 0.5$ \\
SVM, linear kernel     & $0.617 \pm 0.009$ & $5290.1 \pm 205.5$ \\
SVM, polynomial kernel & $0.605 \pm 0.013$ & $10126.3 \pm 413.4$ \\
SVM, Gaussian kernel   & $0.597 \pm 0.039$ & $6331.7 \pm 410.2$ \\
\thickhline
\\[-0.2cm]
\end{tabular}
\caption{{\bf Accuracy and runtime comparisons of RIDDLE and other methods on a 1\% subset of full dataset.}
RIDDLE yielded higher accuracy compared to logistic regression ($p=6.83 \times 10^{-2}$), random forest ($p=1.25 \times 10^{-8}$), support vector machine (SVM) with linear kernel ($p=2.75 \times 10^{-3}$), SVM with polynomial kernel ($p=1.09 \times 10^{-5}$), and SVM with Gaussian kernel ($p=6.43 \times 10^{-2}$). RIDDLE exhibited significantly lower runtimes than the SVM methods (all $p < 10^{-11}$). Scores were averaged over ten $k$-fold cross-validation experiments using a random 1\% subset of the full dataset (150K samples). A standard computing configuration was used: 16 Intel Sandybridge cores at 2.6 GHz; graphics processing units were not utilized.}
\label{table3}
\end{table}
\end{document}